\documentclass[ reprint,nofootinbib,amsmath,amssymb,aps]{revtex4-1}
\usepackage{graphicx}
\usepackage{subfigure}
\usepackage{dcolumn}
\usepackage[colorlinks,linkcolor=magenta,anchorcolor=cyan,citecolor=blue]{hyperref}
\usepackage{bm}
\usepackage[multiple]{footmisc}

\def\be{\begin{equation}}
\def\ee{\end{equation}}
\def\ba{\begin{eqnarray}}
\def\ea{\end{eqnarray}}

\newcommand{\eq}[1]{(\ref{#1})}

\def\q{\theta} \def\r {\rho}         \def\f {\phi}             \def \inf {\infty}  
              \def\grad{\nabla}\def\.{\cdot}
\def\math {\mathcal}
\begin{document}
\title{Dark-Matter Induced Scalarization of Black Holes in Extended Scalar-Tensor-Gauss-Bonnet Theories}
\author{Lei Tang${}^1$}
\author{Jie Jiang${}^2$}
\email{Corresponding author. jiejiang@mail.bnu.edu.cn}
\affiliation{${}^1$College of Physics and Electronic Engineering, Sichuan Normal University, Chengdu 610101, China}
\affiliation{${}^2$Faculty of Arts and Sciences, Beijing Normal University, Zhuhai 519087, China}

\begin{abstract}

In the extended scalar-tensor-Gauss-Bonnet theory, spontaneous scalarization in the GB\(^-\) regime typically occurs only in rotating black holes, while it is absent in spherically symmetric black holes, a phenomenon known as spin-induced scalarization. However, we find that when the spacetime is permeated by perfect fluid dark matter, spontaneous scalarization can also be induced by dark matter in the GB\(^-\) regime. Analytical calculations reveal that this scalarization occurs when the dark matter parameter \(b/M\) exceeds a critical value \((b/M)_\text{crit}\simeq1.86287\), a threshold determined by the lower boundary of the unstable region for scalar perturbations as the coupling constant approaches negative infinity. Additionally, we verified these findings through numerical analysis of the time evolution of scalar perturbations, identifying the unstable parameter region. The results show that when coupling constant \(-\lambda/M^2\) is small, spontaneous scalarization only occurs near the extremal black hole limit. As \(-\lambda/M^2\) increases, the scalarization region expands; however, its lower boundary remains above \(b/M \simeq 1.86287\), consistent with theoretical predictions.

\end{abstract}
\maketitle
\section{Introduction}

The no-hair theorem is a fundamental concept for understanding black hole properties. According to this renowned theorem, an asymptotically flat and stationary black hole spacetime can be uniquely characterized by at most three conserved quantities: mass, angular momentum, and electric charge \cite{Israel:1967wq, Carter:1971zc, Ruffini:1971bza}. This uniqueness implies that matter fields propagating in a stationary spacetime will either be absorbed or scattered by the black hole, eventually decaying and failing to form any persistent features around the black hole. While the no-hair theorem holds robustly within the framework of Einstein's electro-vacuum general relativity (GR), deviations from this theorem have been observed in scenarios beyond GR or in GR with certain non-linear matter sources. For example, in extended gravitational theories, black holes can support additional stable hair through coupling with extra fields, leading to the formation of so-called hairy black holes. This phenomenon has been demonstrated in models such as Einstein-Yang-Mills theory \cite{MSDV, Bizon:1990sr, Greene:1992fw}, dilaton black holes \cite{Kanti:1995vq}, and black holes with Skyrme hair \cite{Luckock:1986tr, Droz:1991cx, Herdeiro:2015waa}. The existence of these hairy black hole solutions not only questions the applicability of the no-hair theorem but also motivates further exploration and expansion of this theory.

In addition to the introduction of extra fields as a way for black holes to acquire additional hair, recent research has shown that spontaneous scalarization is another significant mechanism leading to black holes acquiring scalar hair. Spontaneous scalarization was initially proposed in studies of neutron stars within scalar-tensor theories, where scalarization arises from a tachyonic instability induced by a non-minimal coupling between the scalar field and Ricci curvature \cite{Damour:1993hw}. In recent studies, it has been found that by introducing non-minimal coupling terms between the scalar field and curvature invariants or electromagnetic invariants within extended scalar-tensor theories, black holes can also undergo spontaneous scalarization triggered by tachyonic instability, resulting in scalar hair \cite{Doneva:2021dqn,East:2021bqk,Doneva:2017bvd,Silva:2017uqg,Antoniou:2017acq,Cunha:2019dwb,Herdeiro:2018wub,Doneva:2017bvd,Silva:2017uqg,Doneva:2018rou,Antoniou:2017acq,Cunha:2019dwb,Herdeiro:2020wei,Berti:2020kgk,Dima:2020yac,Bakopoulos:2018nui,Hod:2022hfm,Doneva:2021dcc,Gao:2018acg,Myung:2020etf,Zou:2021ybk,Zhang:2021btn}.

The extended scalar-tensor-Gauss-Bonnet (ESTGB) theory provides a significant framework for studying spontaneous scalarization. In this theory, a direct coupling term \( \lambda \varphi^2 \mathcal{G} \) between the Gauss-Bonnet (GB) curvature invariant \( \mathcal{G} \) and the quadratic term of the scalar field \( \varphi^2 \) is introduced, which can lead to a negative effective mass-squared term in the scalar field perturbation equation, thereby inducing the tachyonic instability and triggering the process of spontaneous scalarization. When the coupling constant \( \lambda > 0 \), both spherically symmetric and rotating black holes in ESTGB theory can undergo spontaneous scalarization, referred to as \( \text{GB}^+ \) spontaneous scalarization \cite{Doneva:2017bvd,Silva:2017uqg,Doneva:2018rou,Cunha:2019dwb}. However, for negative values of the coupling constant, Dima, Barausse, and et.al \cite{Dima:2020yac} found through numerical studies that linear perturbations of non-minimally coupled scalar fields in Kerr black holes lead to instability only when the black hole’s angular momentum exceeds a critical value, initiating the scalarization process. This phenomenon, known as spin-induced scalarization \cite{Herdeiro:2020wei,Berti:2020kgk,Dima:2020yac}, does not occur in spherically symmetric black holes.

Based on current cosmological and astronomical observations, combined with predictions from the standard cosmological model, our universe is thought to be composed of dark matter (DM), dark energy, and baryonic matter. Observations such as the rotation curves of spiral galaxies \cite{Rubin:1970zza}, the large-scale structure formation of the universe \cite{Davis:1985rj}, and the dynamical stability of disk galaxies \cite{Ostriker:1973uit} indirectly confirm the existence of DM. Studying DM is crucial for testing fundamental assumptions of the standard cosmological model and for better understanding the evolution and structure of the universe. In addition, DM may have significant impacts on spacetime and the properties of matter surrounding celestial bodies and black holes \cite{Barausse:2014tra,Cardoso:2021wlq,Cheung:2021bol,Cardoso:2022whc,Destounis:2022obl}. Recent studies on the superradiant instability and accretion phenomena of axion DM clouds around black holes suggest that DM might even be detectable through gravitational wave signals \cite{Traykova:2021dua,Clough:2019jpm,Bamber:2020bpu}. Perfect fluid dark matter (PFDM) is a novel model that treats DM as a perfect fluid \cite{Rahaman:2010xs,Kiselev:2003ah} and has been successfully used to explain the rotation curves of spiral galaxies \cite{Potapov:2016obe}. Numerous spherically symmetric black hole solutions surrounded by PFDM have been obtained \cite{Kiselev:2003ah,Kiselev:2004vy,Kiselev:2004py,Li:2012zx}, with PFDM producing important modifications to the geometry of spacetime in these solutions. An important unresolved question is whether, within the extended scalar-tensor-Gauss-Bonnet (ESTGB) theory with a negative coupling constant, PFDM could influence the spontaneous scalarization of black holes.

This paper aims to investigate the effects of DM on tachyonic instability and spontaneous scalarization of scalar field perturbations in the ESTGB theory. The paper is structured as follows: In Sec. \ref{sec2}, we introduce the theoretical properties of ESTGB theory with PFDM and the spacetime geometry of a PFDM-Schwarzschild black hole when the scalar field vanishes. In Sec. \ref{sec3}, we analyze the field equation for scalar field perturbations in this black hole geometry. Using the necessary condition for tachyonic instability, specifically the presence of a negative effective mass term, we analytically derive the critical condition for the onset of linear spontaneous scalarization. In Sec. \ref{sec4}, we perform a detailed numerical analysis of the time-domain profile of the scalar field, identifying the parameter regions in which the black hole exhibits tachyonic instability and comparing these results with the analytical predictions. Finally, we discuss and summarize the findings of this study.

\section{Description of the system}\label{sec2}

In this section, we will consider the ESTGB theory surrounded by PFDM, which can be described by the action \cite{Doneva:2017bvd,Silva:2017uqg,Li:2012zx,Kiselev:2004vy}
\begin{equation}\label{act}
S=\int d^4 x \sqrt{-g}\left[R-2 \nabla_a \varphi \nabla^a \varphi+F(\varphi) \math{G}+\math{L}_\text{DM}\right],
\end{equation}
where $R$ is the Ricci scalar, $\varphi$ is a real scalar field, $\math{L}_\text{DM}$ gives the Lagrangian density of PFDM, $f(\varphi)$ is some coupling function controlling the  non-minimal coupling between the scalar field and the GB invariant which explicitly reads
\begin{equation}
\math{G}=R_{abcd} R^{abcd}-4 R_{ab} R^{ab}+R^2.
\end{equation}
The field equations of the theory are given by
\ba\begin{aligned}\label{EOM}
R_{ab}-\frac{1}{2}R g_{ab}+\Gamma_{ab}&=T^\text{SC}_{ab}+T^\text{DM}_{ab},\\
\nabla_{a}\nabla^{a}\varphi&=-\frac{\lambda}{4}F'(\varphi)\mathcal{G},
\end{aligned}\ea
with
\ba\begin{aligned}
T^\text{SC}_{ab}=&2\nabla_{a}\varphi\nabla_{b}\varphi-g_{ab}\nabla_{c}\varphi\nabla^{c}\varphi\\
\Gamma_{ab}= &-R(\nabla_{a}\Psi_{b}+\nabla_{b}\Psi_{a})-4\nabla^{c}\Psi_{c}\bigg(R_{ab}-\frac{1}{2}Rg_{ab}\bigg)  \\
&+4R_{ac}\nabla^{c}\Psi_{b}+4R_{bc}\nabla^{c}\Psi_{a}-4g_{ab}R^{cd}\nabla_{c}\Psi_{d}\\
&+4R^{d}{}_{acb}\nabla^{c}\Psi_{d},
\end{aligned}\quad\ea
and the stress-energy tensor $T_{ab}^\text{DM}$ of PFDM. Here we denote
\ba\begin{aligned}
\Psi_a = \frac{dF(\varphi)}{d\varphi} \grad_a\varphi\,.
\end{aligned}\ea
The coupling function $F(\varphi)$ determines the strength of the non-minimal coupling of $\varphi$ to the GB term. To study the scalar field perturbation under first-order approximation, we require $F'(0)=0$, which can be achieved using a quadratic coupling function $1+2\lambda \varphi^2$, an exponential coupling function $e^{2\lambda \varphi^2}$, or other forms that satisfy $F(0)=1$ and $F''(0)=4\lambda$, i.e.,
\ba
F(\varphi)=1 + 2\lambda \varphi^2+O(\varphi^4)\,,
\ea
where $\lambda$ is the coupling constant, which controls the strength of the nontrivial direct interaction between the GB curvature invariant and the nonminimally coupled massless scalar field.

This fundamental coupling parameter, with the dimensions of length squared, can take either positive or negative values. When \( \lambda > 0 \), numerous studies have shown that, even in the absence of DM, a sufficiently large coupling constant \( \lambda \) causes the curvature invariant term to induce tachyonic instability in the scalar field perturbations, leading to the spontaneous scalarization of black holes. This phenomenon is known as \( \text{GB}^+ \) scalarization \cite{Doneva:2017bvd,Silva:2017uqg,Doneva:2018rou,Cunha:2019dwb}. Conversely, when \( \lambda < 0 \), previous research indicates that spontaneous scalarization occurs only if the angular momentum of black hole exceeds a certain critical value, a process referred to as spin-induced scalarization or \( \text{GB}^- \) scalarization \cite{Herdeiro:2020wei,Berti:2020kgk,Dima:2020yac}. In other words, within the ESTGB theory, spontaneous scalarization for spherically symmetric black holes without DM is only possible when \( \lambda > 0 \). If \( \lambda > 0 \), it is plausible that all DM black holes could also exist spontaneous scalarization. This paper, therefore, focuses on the case where \( \lambda < 0 \), investigating whether DM can induce tachyonic instability and spontaneous scalarization in PFDM-ETSGB theory, and exploring the parameter space that permits such scalarization.

In the limit \(\varphi \to 0\), the PFDM-Schwarzschild black hole is a stationary spacetime solution of the PFDM-ESTGB theory, with its line element given by by\cite{Li:2012zx,Kiselev:2004vy}
\be\begin{aligned}\label{metric}
ds^2&=-f(r)dt^2+\frac{1}{f(r)}dr^2+r^2 (d\theta^2+\sin^2 \theta d\phi^2)
\end{aligned}\ee
with the blackening factor
\be\begin{aligned}\label{blackening1}
f(r)=1-\frac{2M}{r}-\frac{b}{r}\text{ln} \left(\frac{r}{|b|}\right)\,,
\end{aligned}\ee
where $M$ represents the mass of the black hole and $b$ is a parameter related to the PFDM density and pressure.

The stress energy-momentum tenor of the PFDM distribution is given by
\ba\begin{aligned}
(T^\text{DM})^{\mu}{}_{\nu}=\text{ diag} (-\r, p_r, p_\q, p_\phi)
\end{aligned}\ea
with
\ba\begin{aligned}\label{rpp}
\r=-p_r=\frac{b}{r^3}\,,\quad\quad p_\q=p_\f=\frac{b}{2r^3}\,.
\end{aligned}\ea
Here, $\rho$ and $p$ represent the energy density and pressure of PFDM, respectively. We impose the weak energy condition $T_{tt}^\text{DM} \geq 0$ outside the event horizon, which requires $b \geq 0$. In this black hole model, the primary focus is on the local effects of DM on the spacetime structure of black hole, without considering the influence of DM as a whole on the asymptotic behavior of spacetime. Therefore, the DM parameter $b$ is only related to the local properties of the black hole and is independent of the overall properties of the universe, such as the DM abundance in the cosmos. Generally, for specific astronomical black holes, the DM parameter $b$ can be constrained by calculating observable quantities such as galaxy rotation curves and gravitational lensing effects, and comparing these with observational data.

\begin{figure}
    \centering
\includegraphics[width=0.48\textwidth]{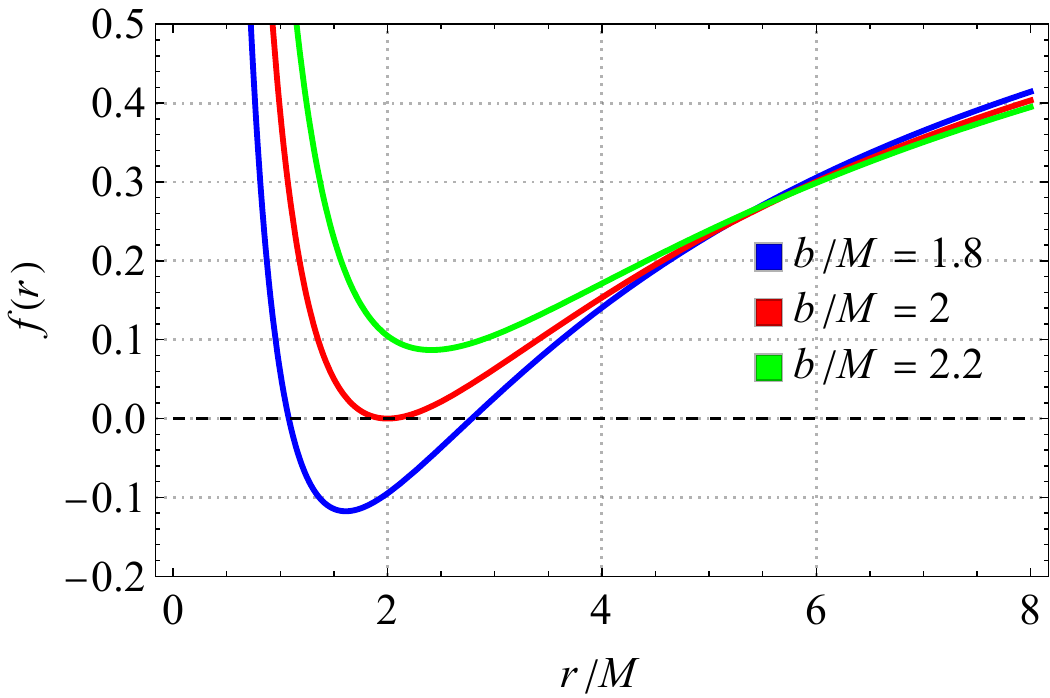}
    \caption{The blackening factor $f(r)$ as a function of $r/M$ with different DM parameter $b/M$.}
    \label{fig0}
\end{figure}

Notably, in this DM black hole model, when \( b > 0 \), a Cauchy horizon (an inner horizon) may exist even if the black hole is uncharged, as shown in Fig. \ref{fig0}. Using the Kruskal coordinates defined in Ref. \cite{Hollands:2019whz}, we can demonstrate that this PFDM-Schwarzschild black hole has the same spacetime structure as a Reissner-Nordström black hole. By setting \( f(r) = 0 \), we obtain the radii of the Cauchy and event horizons, denoted by \( r_- \) and \( r_+ \), respectively, with \( r_- < r_+ \). The extremal black hole is defined when the Cauchy and event horizons coincide, meaning \( f(r_+) = f'(r_+) = 0 \), and at this point, \( b = b_{\text{max}} = 2M \). Therefore, all black hole solutions require the parameter to satisfy \( b/M \leq 2 \).

\begin{figure}
    \centering
\includegraphics[width=0.48\textwidth]{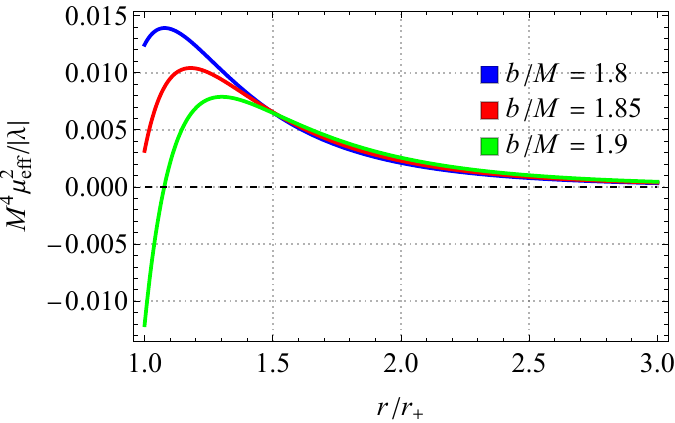}
    \caption{The effective mass term, \(-{M^4 \mu_\text{eff}^2}/{\lambda}\), as a function of \(r/r_+ \in [1, 3]\) for different DM parameters \(b/M = 1.8\), \(1.85\), and \(1.9\) with $\lambda < 0$.}
    \label{fig1}
\end{figure}

\section{Tachyonic instability under the scalar perturbation and onset of scalarization}\label{sec3}

From the field equations \eq{EOM}, the perturbation field equation of the scalar field is given
\begin{equation}\label{kgeq}
\left(\nabla^a \nabla_a-\mu_{\mathrm{eff}}^2\right) \varphi=0,
\end{equation}
in which the effective mass is given by
\ba\begin{aligned}\label{effmass}
\mu_{\mathrm{eff}}^2=-\lambda \math{G}\,.
\end{aligned}\ea
Using the line element \eq{metric} of the PFDM-Schwarzschild black hole solution, we can get
\ba\begin{aligned}\label{mueff}
\mu_\text{eff}^2 = &-\frac{4\lambda}{r^6} \left[ b^2 - 10 b M + 12 M^2  \right.\\
&+ b \ln \left(r/b \right) \left[12 M -5 b  + 3 b \ln \left(r/b \right) \right]\,.
\end{aligned}\ea

Considering the symmetries of the spacetime, we can expand the scalar field as
\be\begin{aligned}\label{sepvar}
\varphi(t,r,\theta,\phi)=\sum_{l,m}Y_{lm}(\theta,\phi)\frac{\psi(r, t)}{r}
\end{aligned}\ee
with the spherical harmonics  $Y_{lm}(\theta,\phi)$. Based on this expansion, the field equation of the scalar perturbation can be expressed as the following Klein-Gordon (KG) equation
\be\begin{aligned}\label{schr}
\frac{\partial^2 \psi(r, t)}{\partial r_*^2}-\frac{\partial^2 \psi(r, t)}{\partial t^2}-V(r)\psi(r)=0
\end{aligned}\ee
with
\be\begin{aligned}\label{potential}
V(r)=&f(r)\left[\frac{l(l+1)}{r^2}+\frac{f'(r)}{r}+\mu_\text{eff}^2\right]\,.
\end{aligned}\ee
Here we used the tortoise coordinate defined by
\be\begin{aligned}
d r_*=\frac{dr}{f(r)}\,.
\end{aligned}\ee

\begin{figure*}
    \centering
\includegraphics[width=0.45\textwidth]{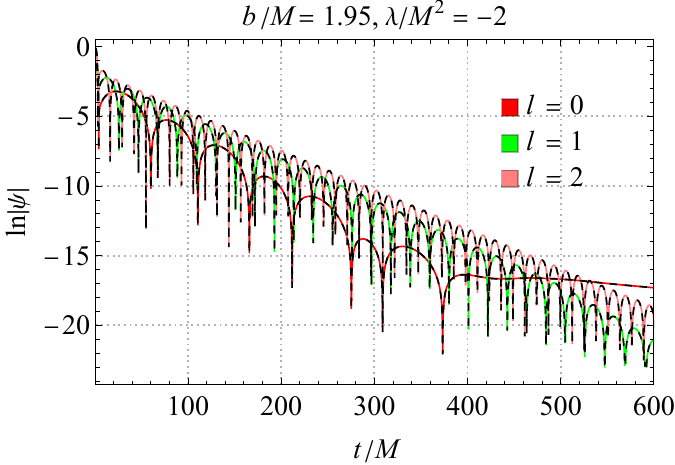}
\includegraphics[width=0.45\textwidth]{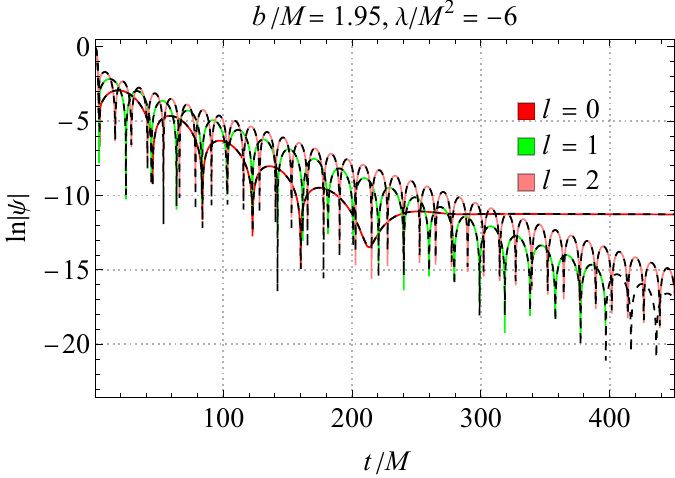}
\includegraphics[width=0.45\textwidth]{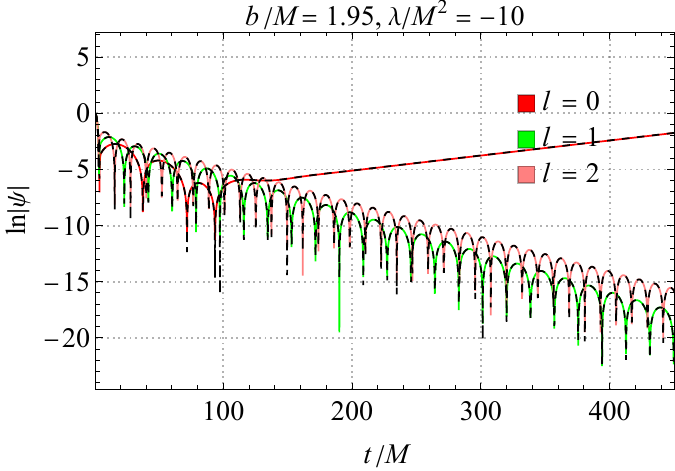}
\includegraphics[width=0.45\textwidth]{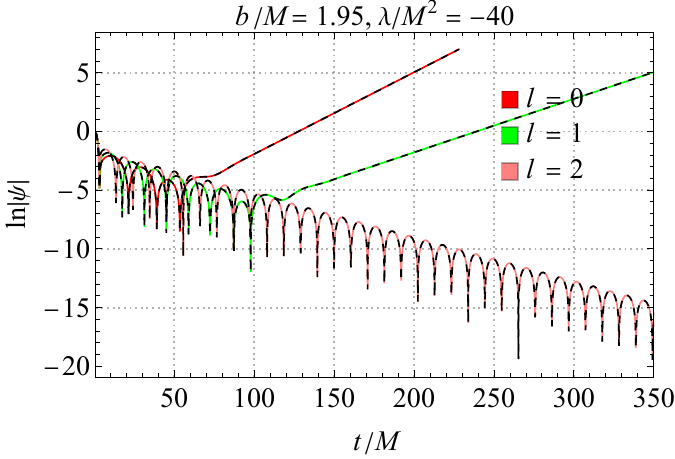}
    \caption{The time evolution of the scalar field perturbation for a given DM parameter \( b/M = 1.95 \) is obtained using the RF method (solid lines) and the FF method (dashed lines) for different coupling constants \(\lambda/M^2\) and angular quantum number \( l \).}
    \label{fig2}
\end{figure*}

A necessary condition for the occurrence of tachyonic instability and the formation of linearly spontaneous scalarized structures is the presence of an attractive (negative) effective potential well in the KG equation \eq{schr} \cite{Cunha:2019dwb,Hod:2020jjy,Hod:2022txa,Hod:2022hfm,Hod:2020jjy}. From the expression of the effective potential \eq{potential}, it is evident that the first two terms in the PFDM-Schwarzschild black hole are always positive. Thus, the primary source of the attractive effective potential that induces spontaneous scalarization in the black hole is the effective mass term, which also reflects the tachyonic instability in the scalar field perturbations. Therefore, a necessary condition for spontaneous scalarization in the spacetime we are considering is the existence of regions outside the event horizon where the effective mass term \( \mu_\text{eff}^2 \) becomes negative.

From the expression \eq{mueff} of the effective mass, it is easy to see
\ba\begin{aligned}
\lim_{r\to\inf} \mu_\text{eff}^2\simeq -\frac{12\lambda b^2 \ln^2 \left(r/b \right)}{r^6}\,.
\end{aligned}\ea
This implies that when \( \lambda > 0 \) and \( r \) is sufficiently large, the effective mass term will always be less than zero. This result indicates that as long as the coupling constant sufficiently large, the spacetime might exist tachyonic instability as well as spontaneous scalarization.

In the following, we consider the situation with the negative coupling constant, i.e., $\lambda < 0$. In this case, we can see that
\ba\begin{aligned}
\mu_\text{eff}^2 = &-\frac{48\lambda M^2}{r^6} > 0
\end{aligned}\ea
without the PFDM, which implies that the scalarization cannot be occurred in the Schwarzschild black hole for ESTGB theory. However, in the presence of PFDM with a negative coupling constant, as shown in Fig. \ref{fig1}, increasing the DM parameter \( b/M \) allows the effective mass term \( \mu_\text{eff}^2 \) of the scalar field perturbations to become negative outside event horizon. This suggests that spontaneous scalarization may still occur under these conditions.

Next, we will use analytical techniques to determine the critical onset line for spontaneous scalarization in the composed physical system, denoted as \( b = b_{\text{crit}} \). As mentioned earlier, tachyonic instability can occur if and only if there exist regions in spacetime outside the event horizon where the effective mass term \( \mu_\text{eff}^2 \) becomes negative, specifically,
\ba
\min_{r>r_+}\{ \mu_{\text{eff}}^2(r; M, b) \} < 0\,,
\ea
Considering that $\lambda<0$, this implies that
\ba
\min_{r>r_+}\{  \math{F}(r; M, b)\} < 0\,,
\ea
in which we denote
\ba\begin{aligned}\label{eqF}
\math{F} \equiv b^2 - 10 b M + 12 M^2 + b z \left[12 M -5 b  + 3 b z \right]
\end{aligned}\ea
with
\ba\begin{aligned}
z \equiv \ln \left(r/b \right)\,.
\end{aligned}\ea

From Eq. \eq{eqF}, it is easy to find that $\math{F} < 0$ only when
\ba\begin{aligned}
\frac{5 b-12 M-\sqrt{13} b}{6 b} < z < \frac{5 b-12 M+\sqrt{13} b}{6 b}\,.
\end{aligned}\ea
That is to say, in order to ensure that there exists a negative value of $\math{F}(r; M, b)$ outside the event horizon (i.e., $r>r_+$), we need
\ba\begin{aligned}\label{ineq1}
\frac{5 b-12 M+\sqrt{13} b}{6 b}>z_+\equiv \ln \left(r_+/b \right)\,.
\end{aligned}\ea
Using $f(r_+) = 0$, we have
\ba\begin{aligned}
z_+=\frac{r_+}{b}-\frac{2 M}{b}\,.
\end{aligned}\ea
Then, inequality \eq{ineq1} reduces to
\ba\begin{aligned}\label{critcd}
\frac{b}{r_+} > \left(\frac{b}{r_+}\right)_\text{crit} = \frac{5-\sqrt{13}}{2}\,.
\end{aligned}\ea
Using $f(r_+) = 0$, we can get
\ba\begin{aligned}
\frac{b}{M} = \frac{2b/r_+}{1+(b/r_+)\ln(b/r_+)}\,.
\end{aligned}\ea
It is easy to check that the left function of the above identity increases with $b/r_+$ when $0 < b/r_+ < 1$. Together with the condition \eq{critcd}, i.e.,  $b/r_+ > (5-\sqrt{13})/2$, the inequality \eq{critcd} reduces to
\ba\begin{aligned}
\frac{b}{M} &> \left(\frac{b}{M}\right)_\text{crit}\\
&= \frac{12}{5+\sqrt{13}-6 \ln [(5+\sqrt{13})/6]}\\
&\simeq 1.86287\,.
\end{aligned}\ea
Therefore, the above discussion suggests that in the PFDM-ESTGB theory with a negative coupling constant, spontaneous scalarization of the spacetime can only occur if the DM parameter \( b/M \) exceeds its critical value \( (b/M)_\text{crit} \approx 1.86287 \). Thus, we refer to this scalarization process as DM induced scalarization.

\section{Numerical calculation of tachyonic instability}\label{sec4}

In this section, we calculate the time-domain profile of scalar field perturbations using numerical methods to analyze their stability and compare these numerical results with the analytical findings from the previous section. There are several approaches to solving the Klein-Gordon equation (\ref{schr}). Here, we primarily use the RF method for calculation, supplemented by the FF method for validation. Specifically, in the RF method, we use the fourth-order Runge-Kutta method for time evolution, while the spatial coordinate \( r_* \) is discretized using the finite difference method. In the FF method, both time and spatial coordinates are discretized using the finite difference method.

\begin{figure*}
    \centering
\includegraphics[width=0.48\textwidth]{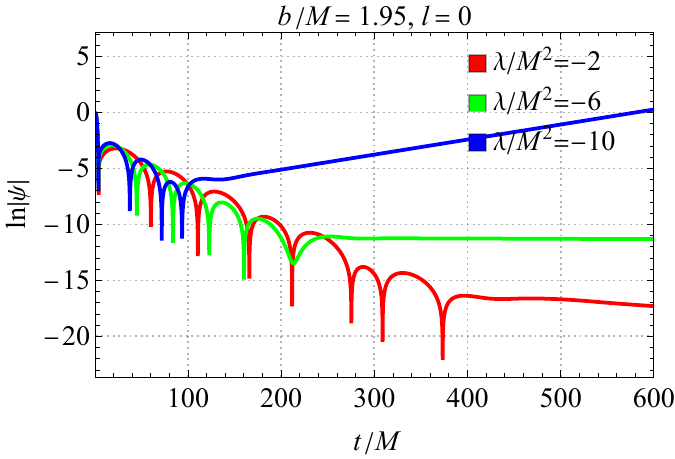}
\includegraphics[width=0.48\textwidth]{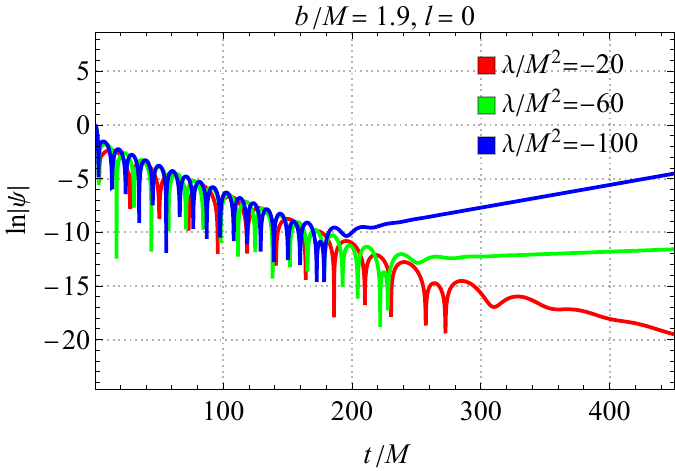}
    \caption{The figures show the time-domain profile at \( l = 0 \) of the scalar perturbation for different coupling constants \(\lambda/M^2\) and DM parameters \( b/M = 1.9 \) (right panel) and \( b/M = 1.95 \) (left panel).}
    \label{fig3}
\end{figure*}

\begin{figure*}
    \centering
\includegraphics[width=0.5\textwidth]{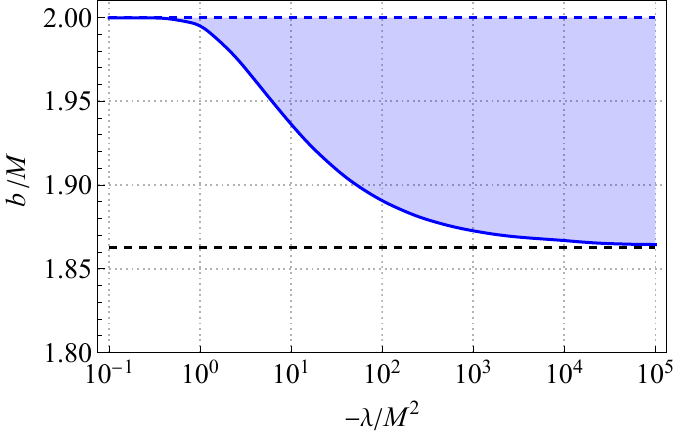}
    \caption{In the \((-\Lambda/M^2) - b/M\) parameter space, the blue shaded region represents the area where scalar field perturbations become unstable, resulting in spontaneous scalarization. The blue curve below outlines the critical condition for spontaneous scalarization at a given coupling constant \(\lambda/M^2\). The upper blue dashed line \((b/M = 2)\) corresponds to the extremal black hole, while the lower black dashed line marks the critical value \(b/M \simeq 1.86287\), which is the threshold necessary for tachyonic instability.}
    \label{fig4}
\end{figure*}

\subsection{Numerical method}

In applying the RF method, we use different computational approaches for the time and spatial coordinates: the fourth-order Runge-Kutta method for time and the finite difference method for space. With the grid indices, \( n \) represents the time step, and \( j \) represents the spatial step, allowing us to precisely describe the behavior of the scalar field \(\psi\) across both dimensions.

For the second-order derivative of $\psi$ with respect to the radial coordinate \( r_* \), the finite difference method provides the following discretized form
\ba\label{Fmethod}
\frac{\partial^2 \psi}{\partial r_*^2} \Big|_{j} \approx \frac{\psi_{j+1} - 2 \psi_{j} + \psi_{j-1}}{(\Delta r_*)^2}\,,
\ea
where \(\Delta r_*\) is the spatial step size. This formulation converts the second-order spatial derivative into an algebraic expression over neighboring grid points, simplifying the numerical calculation.

For the fourth-order Runge-Kutta method in the time coordinate direction, we first need to rewrite the original second-order partial differential equation as two coupled first-order equations. Introducing the variable \(\Pi = {\partial \psi}/{\partial t}\), we decompose the equation as follows
\ba\begin{aligned}
\frac{\partial \psi_j}{\partial t}&= \Pi_{j}\,,\\
\frac{\partial \Pi_j}{\partial t} &= F(\psi_{j-1},\psi_{j},\psi_{j+1})\\
&\equiv\frac{\psi_{j+1} - 2 \psi_{j} + \psi_{j-1}}{(\Delta r_*)^2}-V_j \psi_j.
\end{aligned}\ea
For each time step \(\Delta t\), we compute the following slopes at each spatial point \(j\)
\ba\begin{aligned}
k_{1,j} &= \Delta t \, \Pi_{j}, \quad\,\,\,\, l_{1,j} = \Delta t \, F(\psi_{j-1}^{(n)}, \psi_{j}^{(n)}, \psi_{j+1}^{(n)})\,,\\
k_{2,j} &= \Delta t \, \Pi_{1,j}, \quad l_{2,j} =  \Delta t \, F(\psi_{1,j-1}^{(n)}, \psi_{1,j}^{(n)}, \psi_{1,j+1}^{(n)})\,,\\
k_{3,j} &= \Delta t \,\Pi_{2,j}, \quad l_{3,j} =  \Delta t \, F(\psi_{2,j-1}^{(n)}, \psi_{2,j}^{(n)}, \psi_{2,j+1}^{(n)})\,,\\
k_{4,j} &= \Delta t \, \Pi_{3,j}, \quad l_{4,j} =  \Delta t \, F(\psi_{3,j-1}^{(n)}, \psi_{3,j}^{(n)}, \psi_{3,j+1}^{(n)})
\end{aligned}\ea
with
\ba\begin{aligned}
\psi_{1\,\text{or}\,2,j}^{(n)}=\psi_{j}^{(n)}+\frac{k_{1\,\text{or}\,2,j}}{2}\,,\quad\psi_{3,j}^{(n)}=\psi_{j}^{(n)}+k_{3,j},\\
\Pi_{1\,\text{or}\,2,j}^{(n)}=\Pi_{j}^{(n)}+\frac{l_{1\,\text{or}\,2,j}}{2}\,,\quad\Pi_{3,j}^{(n)}=\Pi_{j}^{(n)}+l_{3,j}\,.\\
\end{aligned}\ea
Then, the updated values of \(\psi\) and \(\Pi\) at each spatial point \(j\) are given by
\ba\begin{aligned}
\psi_{j}^{(n+1)} &= \psi_{j}^{(n)} + \frac{1}{6} (k_{1,j} + 2k_{2,j} + 2k_{3,j} + k_{4,j})\,,\\
\Pi_{j}^{(n+1)} &= \Pi_{j}^{(n)} + \frac{1}{6} (l_{1,j} + 2l_{2,j} + 2l_{3,j} + l_{4,j})\,.
\end{aligned}\ea

For the FF method, both the spatial coordinate and time evolution use the finite difference method. Specifically, spatial discretization still follows Eq. \ref{Fmethod}, and for the time evolution, we have
\ba\begin{aligned}
\psi_j^{(n+1)}&=\psi_j^{(n)}+\Pi_j^{(n)}\Delta t\,,\\ \Pi_j^{(n+1)}&=\Pi_j^{(n)}+F(\psi_{j-1},\psi_{j},\psi_{j+1})\Delta t\,.
\end{aligned}\ea

Using the above formulas, we can apply an iterative method to calculate the time-domain profile of the scalar perturbation. Without loss of generality, the initial condition is chosen as a Gaussian wave packet, defined by
\ba
\psi(0, r_*) = e^{-\frac{(r_* - r_*^c)^2}{2}}, \quad \Pi(0, r_*) = 0,
\ea
where \( r^c_* \) represents the central position of the Gaussian wave packet.

\subsection{Numerical results}

In Fig. \ref{fig2}, we present the time evolution of the scalar perturbation wave function \(\psi(r_*, t)\) at \(r_* = 0\) for a DM parameter \( b/M = 1.95 \), calculated using the FR method (solid lines) and the FF method (dashed lines) for different coupling constants \(\lambda/M^2\) and angular quantum numbers \( l \). Here, \(r_* = 0\) corresponds to the maximum point of the effective potential \( V(r) \). Our results demonstrate consistency between the two methods. Additionally, we observe that, across various angular quantum numbers \( l \), the mode with \( l = 0 \) exhibits the slowest decay or the fastest divergence. Therefore, in subsequent discussions on the stability of scalar field perturbations, we primarily focus on calculations with \( l = 0 \).

To examine whether tachyonic instability indeed exists in the evolution of the scalar field for \( b/M > (b/M)_\text{crit} \simeq 1.86287 M \), Fig. \ref{fig3} illustrates the time-domain profile of the dominant mode (\( l = 0 \)) of the scalar perturbation as the coupling constant \( -\lambda/M^2 \) increases for given DM parameters \( b/M = 1.9 \) and \( 1.95 \). Our results show that for both values of the DM parameter, as the coupling constant \(\lambda/M^2\) increases, the perturbative scalar field becomes increasingly unstable, ultimately leading to spontaneous scalarization of the spacetime.

Finally, by sampling points in the \((-\lambda/M^2) - b/M\) parameter space and calculating the time evolution of the scalar perturbation wave function to assess stability, Fig. \ref{fig4} illustrates the region in which scalar field perturbations become unstable in \((-\lambda/M^2) - b/M\) parameter space. This indicates where spontaneous scalarization occurs for PFDM black holes in the PFDM-ESTGB theory. The blue shading in the figure represents the critical conditions for spontaneous scalarization. From this figure, it is clear that when the coupling constant \(-\lambda/M^2\) is small, spontaneous scalarization can only occur in regions very close to the extremal black hole limit \(b/M \to (b/M)_\text{max} = 2\). As \(-\lambda/M^2\) increases, the parameter range for spontaneous scalarization broadens. However, its lower boundary does not exceed the critical line \(b/M = (b/M)_\text{crit}\simeq 1.86287\) derived in the previous section. This implies that the critical line \(b/M \simeq 1.86287\) represents the threshold for spontaneous scalarization as \(-\lambda/M^2\) approaches infinity.

\section{Discussion and Conclusion}\label{sec5}

In the ESTGB theory, it is well known that spontaneous scalarization in the GB\(^-\) regime (i.e., \(\lambda < 0\)) does not occur in spherical black holes because the effective mass term \(\mu_\text{eff}^2\) is always positive outside the event horizon. The existence of DM in our universe has been indirectly confirmed by numerous astronomical observations \cite{Rubin:1970zza,Davis:1985rj,Ostriker:1973uit}. As a model for DM, the PFDM model can explain these astronomical phenomena \cite{Rahaman:2010xs,Kiselev:2003ah}. Interestingly, by imposing the weak energy condition on DM (specifically, requiring \(b > 0\)), our study reveals that when the DM parameter exceeds a certain critical value,
\ba\begin{aligned}
\left(\frac{b}{M}\right)_\text{crit}= \frac{12}{5+\sqrt{13}-6 \ln [(5+\sqrt{13})/6]}\simeq 1.86287,\quad\quad
\end{aligned}\ea
PFDM can induce tachyonic instability in the scalar field perturbations even in the GB\(^-\) regime, leading to spontaneous scalarization of the black hole. We refer to this as DM-induced spontaneous scalarization.

First, based on the necessary condition for spontaneous scalarization—that there must exist regions outside the event horizon where the effective mass term \(\mu_\text{eff}^2\) becomes negative—we analytically calculated the critical DM parameter required for spontaneous scalarization. This represents the minimum threshold to induce instability (see Fig. \ref{fig4}). Then, to study tachyonic instability (linear spontaneous scalarization) for a given coupling constant \(\lambda/M^2\), we used numerical calculations to analyze the time-domain profile of scalar perturbations and identified the parameter regions exhibiting instability in the \((-\lambda/M^2) - b/M\) space. The results indicate that when the coupling constant \(-\lambda/M^2\) is small, spontaneous scalarization can only occur in regions very close to the extremal black hole limit \(b/M \to (b/M)_\text{max} = 2\). As \(-\lambda/M^2\) increases, the parameter range for spontaneous scalarization expands, but its lower boundary does not exceed the critical line \((b/M)_\text{crit} \simeq 1.86287\), consistent with the analytical results obtained.

Moreover, it is worth mentioning that our results show the relationship between the reduced DM parameter \( b/M \) and tachyonic instability of black hole, which indicates that regardless of the constraints on the DM parameter \( b \), as long as the mass \( M \) is sufficiently small such that \( b/M \) exceeds the critical value, the black hole will induce tachyonic instability, thereby triggering spontaneous scalarization. Therefore, our results also imply that spontaneous scalarization can only occur when the mass of black hole and the DM parameter are of the same order of magnitude.

In our paper, we only consider the tachyonic instability of the scalar field in the PFDM-ESTGB theory. However, the tachyonic instability is merely a necessary condition for the formation of scalarized black holes and does not guarantee the existence of stable branches of scalarized solutions. Therefore, strictly speaking, to fully understand the impact of PFDM on the spontaneous scalarization of black holes, it is necessary to solve the Einstein equations with PFDM to obtain scalarized black hole solutions and analyze their stability.

\section*{Acknowledgments}

J. J. is supported by the National Natural Science Foundation of China with Grant No. 12205014, the
Guangdong Basic and Applied Research Foundation with Grant No. 2021A1515110913.

\end{document}